# Magneto-mechanical reservoir computing combining a two-dimensional network of nonlinear mass-spring resonators with magnetic tunnel junctions

A. Grimaldi, *Member, IEEE*, D. R. Rodrigues, *Member, IEEE*, A. Meo, *Member, IEEE*, F. Garescì, and G. Finocchio, *Senior Member, IEEE*

*Abstract*— Coupled networks of mass–spring resonators have attracted growing attention across multiple fundamental and applied research directions, including reservoir computing for artificial intelligence. This has led to the exploration of platforms capable of tasks such as acoustic-wave classification, smart sensing, predictive maintenance, and adaptive vibration control.

This work introduces a multiphysics reservoir based on a two-dimensional network of coupled nonlinear mass-spring resonators. Each mass has a magnetic tunnel junction on top of it, working as spin diode, used as a spintronic read-out.

As a proof-of-concept, we have benchmarked this reservoir with the task of vowel-recognition reaching accuracy above 95%. Because the device accepts elastic excitations directly, signal injections are simplified, making it well suited for real-time sensing and edge computation. We also studied the effect of nonlinearity, demonstrating how it influences the reservoir dynamics, and assessed its robustness under node-to-node variation of the elastic constants.

## I. Introduction

Recent years have seen a rising interest in mass-spring resonators due to their rich dynamics and their massive use in Information and Communications Technology as building blocks of accelerometers and gyroscopes. When connected as chains, it is possible to manipulate wave propagation and vibrations by tuning their properties, making them also suitable for fundamental studies and applications [1], like vibration isolation [2] and acoustic cloaking [3]. The recent advancements in artificial intelligence and machine learning, together with the ever-increasing energy consumption of training and inference of the models used in such applications, have led to researches tailored to develop neuromorphic computing blocks directly with physical systems. Among the many neuromorphic computing paradigms at the state-of-the-art, reservoir computing (RC) is compatible with nonlinear mechanical resonators [4], promising energetically efficient computation and high scalability potential [5–7].

In this work, built on top of [8], we proposed and benchmarked a multiphysics RC combining coupled nonlinear mass-spring resonators (Duffing oscillators) with spintronic technology used as an effective read-out mechanism. The resonators are organized in a two-dimensional network. The spintronic device used is based on magnetic tunnel junctions (MTJs) working as injection locked spin diodes. This scheme is designed in such a way that the MTJs are magnetically coupled via the stray field but do not influence the mechanical dynamics, i.e. the magnetic force is negligible compared to the elastic force. This approach enables a precise electrical detection of nano-displacements by a direct change of the rectified voltage of the MTJ driven by the displacements induced in the masses of the resonator network [9,10]. We demonstrate that this reservoir coupled with a simple classification layer can perform temporal classification tasks through a vowel recognition benchmark. Our system achieves over 95% accuracy, matching the performance of state-of-the-art RC implementations [11].

The system's performance has also been studied as a function of the nonlinearity and the effect of inhomogeneities of the elastic properties. A key result is that the performance of the RC does not deteriorate showing robustness on spring-to-spring elastic constant variation. This magneto-mechanical reservoir is designed to work with the direct detection of vibrations or acoustic waves with an integrated electrical readout that simplifies the interface between analog physical input and digital computational blocks, making this platform particularly suitable for effective edge AI applications such as real-time biosensing, machinery monitoring, predictive maintenance, and voice recognition.

## II. Model of the Magneto-Mechanical RC

### A) Mass-spring resonator network

Fig. 1a shows a sketch of the RC composed by a continuous elastic medium upon which a two-dimensional grid of MTJ-based spin diodes is positioned. It is designed so that the spin diodes are at a distance that allows magnetic coupling among them [9]. To simulate such a system, where it is fundamental to determine the position of each spin diode relative to the other ones, we model it as a two-dimensional network of mass-spring resonators with a Duffing potential (see Fig. 1b). The spin diodes are located on top of each mass, and their magnetic coupling with the other spin diodes results in a distance-dependent rectification voltage ($V_{\text{dc}}$) [9], as shown schematically in Fig. 1c. A single elastic link is modelled by considering a damped spring-mass Duffing oscillator, whose dynamics is described by

$$m\ddot{\boldsymbol{s}} = -\eta\dot{\boldsymbol{s}} - k_{\text{el}}\boldsymbol{s} - \beta\boldsymbol{s}^3 + m\boldsymbol{a}_{\text{ext}}(t)\,, \quad (1)$$

where $m$ is the mass of the mechanical oscillator, $\eta$ is the viscous damping coefficient, $k_{\text{el}}$ is the elastic constant, $\beta$ is the Duffing coefficient, $\boldsymbol{a}_{\text{ext}}(t)$ is the external acceleration, and $\boldsymbol{s}$ is the displacement of the mass from its resting position. This system is designed in such a way that the effective force arising from the magnetic interactions (stray fields) among the MTJs does not influence the dynamics of the mass-spring resonator,

This work was supported by the project PRIN_20225YF2S4 – MMAGYC, funded by the Italian Ministry of University and Research. G. F. is the corresponding author. A.G., D. R. R., and A. M. are with Department of Electrical and Information Engineering, Politecnico di Bari, Bari, Italy. F. G. is with the Department of Engineering, University of Messina, Messina, Italy. G. F. is with the Department of Mathematical and Computer Sciences, Physical Sciences and Earth Sciences, University of Messina, Messina, Italy (email: gfinocchio@unime.it).

this is why it was neglected in Eq. (1). Each spring connects two elements of the grid network of Fig. 1b. The dynamical equations describing the whole system can be written as

$$m_i \ddot{s}_i = \sum_{\langle i,j \rangle \in E} \left( -\eta \Delta \dot{s}_{ij} - k_{el} \Delta s_{ij} - \beta \Delta s_{ij}^3 + m_i a_{i,ext}(t) \right), \quad (2)$$

where $\ddot{s}_i$ is the acceleration of the $i^{th}$ node, $\Delta \dot{s}_{ij}$ and $\Delta s_{ij}$ are the relative velocity and displacement variables of the connected mass $j$ with respect to mass $i$. The sum is over all pairs $\langle i,j \rangle$ that are in the ensemble of edges of the network shown in Fig. 1b. Table I, section elastic network simulation parameters, summarizes the parameters used for the simulations in the next section.

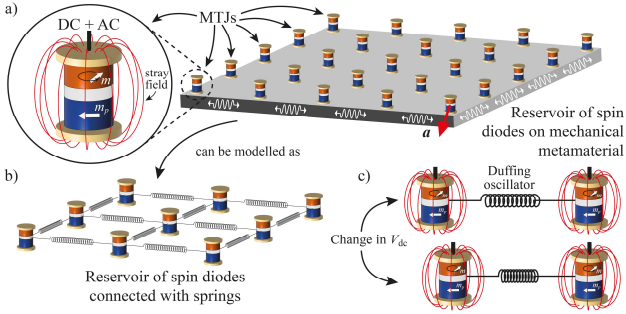

Fig. 1. a) Scheme of the RC. The MTJs are built on top of an elastic medium. The MTJs are hybrid (out-of-plane FL, in-plane PL) and are used as spin diodes, supplied by both direct and alternating currents. b) The system can be modelled as a network of masses on a grid connected with their first neighbors through springs. c) Each spring is treated as a Duffing oscillator. Due to stray field interactions, the rectified voltage $V_{dc}$ across each MTJ depends on its relative distance with the other MTJs of the network.

*B) MTJ-based spin diode*

The MTJ has a free layer (FL) with an out-of-plane easy axis and an in-plane polarizer (PL). The MTJ is biased by both direct and alternating currents in order to work as injection locked spin diode where the normalized magnetization $\boldsymbol{m}$ of the FL of each MTJ oscillates with a self-oscillation frequency $f_0$ given by the dc bias current $I_{dc}$ [12]. From a theoretical point of view, the dynamics of $\boldsymbol{m}$ is described by the Landau-Lifshitz-Gilbert-Slonczewski (LLGS) equation:

$$\frac{d\boldsymbol{m}}{dt} = -\frac{\gamma_0 M_s}{1 + \alpha_G^2} \left( (\boldsymbol{m} \times \boldsymbol{h}_{eff}) + \alpha_G (\boldsymbol{m} \times \boldsymbol{m} \times \boldsymbol{h}_{eff}) \right) + \boldsymbol{T}, \quad (3)$$

where $M_s$ is the saturation magnetization, $\gamma_0$ is the gyromagnetic ratio, $\alpha_G$ is the Gilbert damping, and $\boldsymbol{h}_{eff}$ is the normalized effective field. $\boldsymbol{T}$ is the spin-transfer torque (STT) [9]. The effective field $\boldsymbol{h}_{eff}$ contains the contributions of the uniaxial anisotropy field, $\boldsymbol{h}_u = 2K_u/(\mu_0 M_s^2)(\boldsymbol{m} \cdot \boldsymbol{u}_k)\boldsymbol{u}_k$, the demagnetizing field, $\boldsymbol{h}_{demag} = -N\boldsymbol{m}$, $N$ being the demagnetizing tensor computed by full micromagnetic simulations, and the stray field defined as

$$\sum_{\langle i,j \rangle \in E} \boldsymbol{h}_{stray, j \to i} = -\frac{\mu_0}{4\pi} \sum_{\langle i,j \rangle \in E} \frac{1}{|\Delta s_{ij}|^3} (3\Delta \hat{s}_{ij}(\boldsymbol{m}_j \cdot \Delta \hat{s}_{ij}) + \boldsymbol{m}_j). \quad (4)$$

For additional details about the meaning and the values of the parameters, see Table I section micromagnetic simulations. The STT components contained in $\boldsymbol{T}$ are both the damping-like and field-like torques:

$$\boldsymbol{T} = I_{STT}(t) \frac{g\mu_B}{|e|SM_s l_z} g_T(\boldsymbol{m})(\boldsymbol{m} \times \boldsymbol{m} \times \boldsymbol{p} - q_{STT}\boldsymbol{m} \times \boldsymbol{p}). \quad (5)$$

where $g$ is the Landé factor, $\mu_B$ is the Bohr magneton, $l_z$ and $S$ are, respectively, the thickness and the cross section of the FL, and $g_T = 2P/(1 + P^2 \boldsymbol{m} \cdot \boldsymbol{p})$ is the spin polarization function. $I_{STT}(t)$ is the total current density injected in the MTJ, equal to $I_{STT}(t) = I_{dc} + I_{ac} \sin(2\pi f_{ac} t)$.

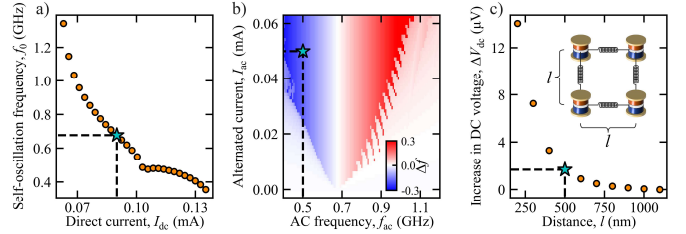

Fig. 2. a) Self-oscillation frequency as a function of $I_{dc}$. The working parameters chosen are highlighted with a star marker and black dashed lines. b) Arnold tongue showing the response of the MTJ for $I_{dc}$= 0.09 mA as a function of the $I_{ac}$ frequency and amplitude. The working parameters chosen are highlighted with a star marker and black dashed lines. c) Variation of the rectified voltage of an isolated device as a function of the relative distance between two MTJ. A scheme of the network is shown in an inset.

TABLE I.    SIMULATION PARAMETERS

| Elastic network simulation | | |
|---|---|---|
| *Parameter name* | *Value* | *Unit* |
| Mass, $m$ | 1 | mg |
| Spring elastic constant, $k_{el}$ | $10^3$ | N/m |
| Viscous damping coefficient, $\eta$ | $10^{-2}$ | Kg/s |
| Duffing coefficient, $\beta$ | See text | N/m$^3$ |
| Grid distance between MTJs, $l$ | 500 | nm |
| External acceleration amplitude, $a_{ext}$ | $9.81 \cdot 10^{-5}$ | m/s$^2$ |
| **Micromagnetic simulation** | | |
| *Parameter name* | *Value* | *Unit* |
| Device size, $l_x \times l_y \times l_z$ | $150 \times 70 \times 1.58$ | nm |
| Demagnetizing tensor $x$, $N_x$ | 0.014 | – |
| Demagnetizing tensor $y$, $N_y$ | 0.040 | – |
| Demagnetizing tensor $z$, $N_z$ | 0.946 | – |
| Saturation magnetization, $M_s$ | $9.5 \cdot 10^5$ | A/m |
| Gilbert damping, $\alpha_G$ | 0.02 | – |
| Uniaxial anisotropy coefficient, $K_u$ | $5.45 \cdot 10^5$ | J/m$^3$ |
| Uniaxial anisotropy easy axis, $\boldsymbol{u}_k$ | $+z$ (out of plane) | – |
| Polarizer layer direction, $\boldsymbol{p}$ | $-x$ (in-plane) | – |
| Direct current, $I_{dc}$ | 0.09 | mA |
| Alternating current amplitude, $I_{ac}$ | 0.05 | mA |
| Alternating current frequency, $f_{ac}$ | 0.5 | GHz |
| Spin polarization coefficient, $P$ | 0.66 | – |
| Field-like torque coefficient, $q_{STT}$ | 0.075 | – |
| MTJ resistances, $R_P - R_{AP}$ | $640 - 1200$ | Ω |

The dependence of the self-oscillation frequency as a function of the $I_{dc}$ computed for a single MTJ is shown in Fig. 2a. In the rest of the paper, we consider a value of $I_{dc}$ equal to 0.09 mA (see the star marker in Fig. 2a), but qualitative similar results are obtained for other $I_{dc}$ values. The MTJ is then supplied by an ac current near $f_0$ to achieve injection-locking [13]. Fig. 2b shows the Arnold tongue obtained for $I_{ac}$ up to 0.06 mA. The chosen $f_{ac}$ and $I_{ac}$ values are marked with dashed black lines meeting in a green star marker. The color represents the difference in oscillation frequency compared to $f_0$. The working parameters ensure that the spin diode is injection-locked similar results have been also obtained for other points of the Arnold tongue.

The grid rest distance between the masses is fixed at $l$ = 500nm but similar results are obtained for 400 and 600 nm. Fig. 2c shows the variation of the rectified voltage $V_{dc}$ across a given MTJ coupled with another one as a function of their distance. The chosen grid distance is indicated with a star marker and black dashed lines.

The simulations were performed within the macrospin approximation with the parameters in Table I. [9]. As shown in Fig. 2b, the spin diodes operate at 500 MHz; the elastic network, on the other hand, is designed to resonate at a much slower frequency, with displacements of few tens of nanometers taking tens of microseconds. This mismatch allows us to approach the simulation of the magnetic system in a quasi-static way, performing a micromagnetic simulation at each mechanical time step while considering the stray field (and the spin diodes positions) fixed for the time required for the magnetization to reach a new stationary state.

*C) Working principle of RC and potential challenges*

RC takes its name from the reservoir, a fixed, often randomly connected recurrent layer. The input of the RC application is fed into this reservoir and mapped into a representation with higher number of dimensions. This high-dimensional output is then fed into a single layer of a neural network to perform the required task. As the reservoir does not need to be trained, in its initial formulation only the output layer is trained using a simple linear regression, eliminating the need for complex and energy-expensive backpropagation operations [14]. Thanks to its compact and energy efficient implementations, RC offers possible applications in real-time data analysis in areas such as automation [15], cyber security [16], medical diagnostics [17] and edge computing [7,18], where low latency and on-device processing are critical.

RC is, by design, compatible with analog and hardware implementations, as physical systems often provide the nonlinear dynamics needed by RC to map the input into higher dimensions. In many cases, this advantage also translates to RC being developed in ultra-compact and low-power computing systems [5,6,19,20]. Physical reservoirs have been demonstrated with many systems, like photonic [21], mechanical [22,23] and magnetic [24] ones. Also other types of mechanical systems have been investigated and shown to work with RC, like optomechanical systems [25] and even paper-based systems [26]. Some of the physical system also introduce challenges that need to be addressed for wider adoption [6,7]. Some of the RC applications with physical systems may be intrinsically limited in scalability or tuneability, while some are bottlenecked by their challenging co-integration with conventional computing architectures. Depending on the system, there might be a mismatch between the timescale of the dynamics of the reservoir and of the computational task considered. Practical issues like limited sampling bandwidth, storage capacity, hardware drift, device-to-device variations, and recalibration requirements also need to be addressed when considering a specific application to make it viable.

*D) RC implementation based on mass-spring resonators*

At nano- and micro-scale, mass-spring resonators can be implemented with piezoelectric materials [7,27,28], that can provide intrinsic nonlinearity and short-term memory, key features for physical RC [29–31]. In addition, their direct strain-induced voltage generation reduce the complexity of the external readout circuitry, reduces its power consumption and enables large-scale integration. Another approach is based on conventional micro-electromechanical systems (MEMS). Although they are usually larger and less sensitive [32], MEMS also exhibit nonlinear resonant dynamics, which can be modelled using the Duffing equation, making them suitable for time-series processing and neuromorphic applications [4,33,34]. Recent advances demonstrate their compatibility with complex computational tasks, including speech classification and real-time sensing [33,35]. Spin diodes have the potential to be integrated with both piezoelectric materials and MEMS, adding functionalities to state-of-the-art mechanical RC paradigms.

### III. PROBLEM SETUP

This RC was benchmarked with a well-known vowel recognition task being already used to evaluate the performance of other RC implementations [11]. Vowel sounds, in phonetics, are usually characterized by characteristic frequencies called formants. To set up vowel recognition, here we use the Hillebrand database [36]. It contains the formant frequencies of vowel sounds spoken by men, women and children. For our application, we use a subset of the database, limiting the data points to male spoken vowels among "ae", "ah", "er", "ih", "iy", "oa", "uw". The data are supplied to this magneto-mechanical RC considering each frequency linearly transformed to fall into the working frequencies of the mass-spring resonators. The full dataset considered, comprising of 243 vowel sounds, is shown in Fig. 3a, with frequencies ranging from 1 to 5 kHz. The two scaled formant frequencies, $f_A$ and $f_B$, are used for the sinusoidal external accelerations acting on the bottom-left and upper-right spin diodes of the 3x3 RC network, sketched on the right and labeled with capital letters A to I. Both accelerations are applied with the same intensity, $\vec{a}_A$ forming a $-120°$ angle with the $x$-axis, and $\vec{a}_B$ a $30°$ angle, respectively. We have selected this configuration because the external masses are more easily accessible. However, qualitative similar results are observed for other choices. Each RC simulation was conducted with a time step of 10 μs, lasting 30 ms in total. An example of elastic response to the scaled formant frequencies of different vowel sounds is shown in Fig. 3b.

The panels are arranged according to the grid structure of the elastic system, the label at the top of each column identifying the $x$ grid coordinate, and each panel showing the oscillations of the $y$ position of the respective spin diode around its $y$ grid coordinate. Each panel also displays the capital letter used to identify each node in Fig. 3a. The bottom-left and upper-right panels refer to the spin diodes to which the external accelerations $\vec{a}_A$ and $\vec{a}_B$ are directly applied. The evolution shown is an extract of the full run (between 18 and 20 ms) to better show the different oscillation patterns each spin diode produces to different vowel sounds.

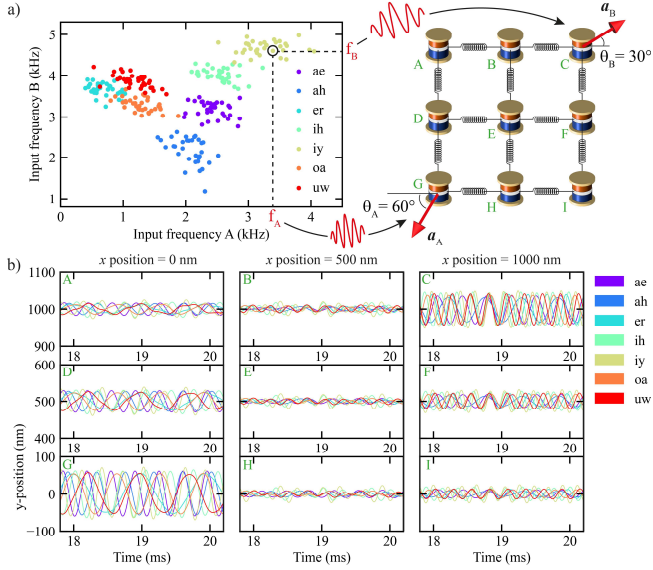

Fig. 3. Input encoding and processing of the vowel recognition problem. a) A vowel input data, expressed, in terms of its formant frequencies, is applied to the network (3 × 3) as sinusoidal external accelerations with slanted angles on opposite corners of the grid with linearly scaled frequencies. b) After a brief transient, the system shows a steady-state dynamics. Each panel shows the $y$-component of the trajectory of a node of the grid for each vowel type considered. The panels are positioned according to the node of the grid they represent, and they are labelled with a letter that matches their respective MTJ in panel (a).

For each mechanical step, a micromagnetic simulation was performed with a time step of 1 ps for a total of 100 ns. The instantaneous rectified voltage response of the spin diode is obtained using Ohm's law, $V(t) = R(t)I(t)$. The current is the total injected current $I(t) = I_{dc} + I_{ac} \sin(2\pi f_{ac} t)$, while the resistance of the MTJ is given by $R(t) = (1 + \hat{p} \cdot m(t))(R_{AP} - R_P)/2$. Where $R_{AP}$ and $R_P$ are respectively the parallel and antiparallel resistance of the MTJ. The average of $V(t)$ across the last 40 ns of the simulation represents the rectified voltage measured at a given step of the mechanical simulation. This is averaged further across the 30 ms of the mechanical simulation, resulting in nine voltage values per vowel sound. Those data are then normalized in a 0-1 range per spin diode, resulting in the final input dataset of the trainable linear combination layer of the RC. Fig. 4a shows the histograms for the normalized voltage values of each vowel for each spin diode of the network in the case in which the spring-mass resonators are treated as linear springs (Duffing parameter $\beta = 0$ N/m³). Each panel is labelled according to the scheme in Fig. 3a. The average normalized voltage value of each histogram is marked with a vertical dashed line. The dataset constituted by the normalized voltage values and their respective vowel sound is used to train a single layer of neural network with SoftMax activation function, the output being the one-hot encoding of the respective vowel category. The data were split into 80% for training and 20% for validation of the accuracy. The loss used is the categorical cross-entropy, as the correct vowel sound is expressed as a one-hot vector. The total number of trainable parameters is $9 \cdot 7 + 7 = 70$. The number of inputs per data point is increased from two (the formants of the vowels) to nine (the measured voltages at each node), effectively increasing the dimensionality of the data. As the number of datapoints is relatively contained, the training was conducted for $10^4$ epochs. This was necessary to train the weights without increasing the learning rate to values that could incur in numerical errors during gradient descent.

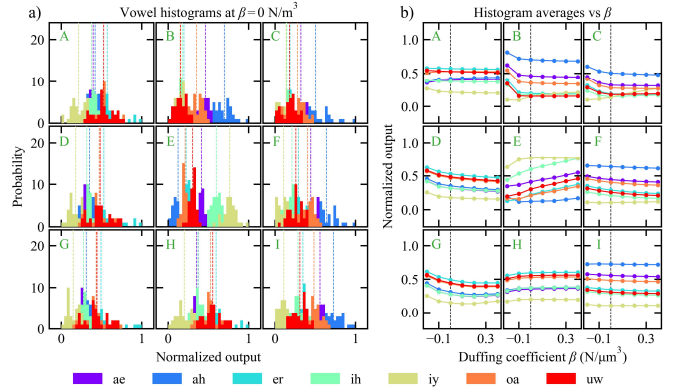

Fig. 4. a) Vowel histograms showing the normalized $V_{dc}$ values of each vowel sound for each spin diode of the elastic network with the Duffing coefficient $\beta = 0$ N/m³. The panels are arranged according to the network grid, as in Fig. 3b, with the letter referring to Fig. 3a. In each panel, the average of the histograms are marked by vertical dashed lines. b) Trend of the histogram averages as a function of the Duffing coefficient $\beta$ for each node of the network. The black vertical dashed line markes the linear spring case, shown in (a).

## IV. RESULTS

The effect of the non-linear Duffing coefficient $\beta$ on the performance of the RC has been investigated for the problem introduced in the previous section. Fig. 4b shows the trend of the average normalized voltage values for each vowel type in each of the nodes of the elastic network as a function of the Duffing coefficient. Negative values of $\beta$ mean that the oscillators act as "soft" springs, with restoring force weaker as the displacement increases. This negative value has a practical limit that depends on the strength of the forces it is subjected to: if the mean range of displacements exceeds $\Delta s_{rms} = \pm\sqrt{-k_{el}/3\beta}$, the spring is too soft for the applied perturbations and it "breaks". In this case, as the perturbations for the linear spring have an approximate mean displacement of 40 nm, the lowest $\beta$ must be no smaller than $-0.2$ N/μm³. Conversely, positive values of $\beta$ cause the oscillators to act as "hard" springs, with the restoring force becoming stronger as the displacement increases. As the effective elastic constant always remains positive in this case, there is no theoretical upper boundary to the Duffing coefficient.

The panels are labelled with the capital letter referring to Fig. 3a. The black dashed line marks the linear spring case, its histograms shown in Fig. 4a. The trends show, as expected, that the Duffing coefficient has more effect on the motion of the masses that are subject to larger displacements. The averages on the non-perturbed corners of the grid remain the same even as $\beta$ changes. The results of the training are shown in Fig. 5a, in which 10 training tests (with randomized validation splits) are performed for the datasets obtained with each $\beta$ value. Fig. 5b shows the final accuracy values along with their standard deviation across the tests. The trend shows that a small positive $\beta$ is the one that results in better performance, reaching about ~87% accuracy.

To study the reliability of this RC design, we investigated deployments where the elastic characteristics of the mass-spring resonators are slightly different than the nominal value. To test the effect that such inhomogeneities have on the performance, we selected a $\beta$ value of 0.2 N/µm³ and randomized the $k_{el}$ of each connection by sampling its value from a Gaussian distribution with mean $\mu_{k_{el}} = 10^3$ N/m and standard deviation $\sigma_{k_{el}} = 10^2$ N/m, meaning that two thirds of the sampled elastic coefficient values have a relative error of less than 10% from the average.

Fig. 5c shows the results of training as a function of the number of epochs, while Fig. 5d shows the average final accuracy and its error among the ten tests performed. In the same panel, the performance of the ideal case with no inhomogeneities is labelled as "Ref". The performance due to non-ideal elastic constant distribution generally improves from the ideal case. In some cases, the magneto-mechanical reservoir performs consistently with over 95% accuracy, with even less variability than the ideal case. This suggests that the non-linearity introduced by these slight inhomogeneities, if they are not negligible, can have a positive effect on separation properties of the RC. This is the proof of concept of the working principle of this multiphysics RC. Large variations between different configurations can be observed, which can be attributed to the small size of the resonator network. We wish to stress that although a simple neural classifier is used, the reservoir itself remains untrained and task-agnostic.

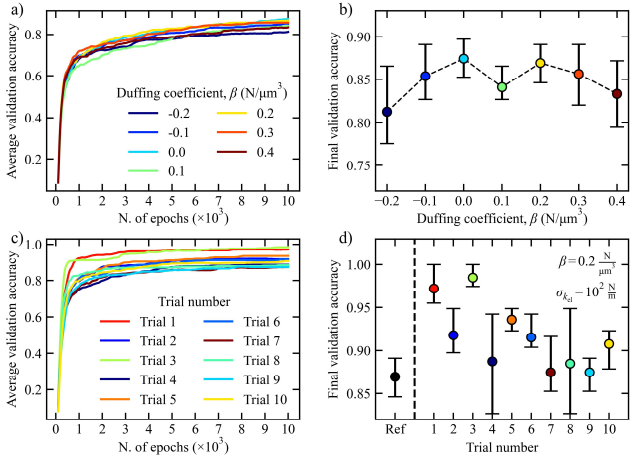

Fig. 5. a) Validation accuracy of the final linear combination of the reservoir computer as a function of the number of epochs for each Duffing coefficient tested. Each curve is averaged across ten tests. b) Final validation accuracy of the tests shown in (a) with the respective standard deviation. c) Validation accuracy of the final linear combination of the reservoir computer as a function of the number of epochs for ten randomized sets of grid parameters. Each curve is averaged across ten tests. d) Final validation accuracy of the tests shown in (c) with the respective standard deviation.

## V. CONCLUSIONS AND OUTLOOK

In this work, we have designed a RC realized with a two-dimensional network of coupled non-linear mass spring resonators having a direct read-out mechanism of the displacements based on spintronic technology. The performance of the RC was benchmarked on a vowel recognition task, where formant frequencies from the well-known Hillebrand database were converted into sinusoidal accelerations applied to different external masses of the elastic network. We studied the performance of such a system as a function of the Duffing coefficient. We also demonstrated the reliability of the RC in presence of a network composed by elastic constants slightly different than their nominal value. In the best case the use of this RC allowed to achieve over 95% accuracy, comparable to other state-of-the-art RC implementations [11]. We emphasize that the performance of the proposed device can improve further by considering a larger grid, which increases nonlinearity and expands the reservoir's capacity. Notably, this system requires no calibration of the spin diodes and only minimal preprocessing limited to a simple linear transformation of the input data into the frequency values within the operating range of the mechanical system. This preprocessing step becomes unnecessary when the input consists of direct vibrations or acoustic waves applied to the mechanical metamaterial.

A key advantage of our proposed device is its high compactness of the reading scheme, as the spin-diode network can be scaled down to the nanometric size while operating across a broad frequency spectrum, from Hz to GHz and can be integrated with piezoelectric materials and MEMS [4] or more complex mechanical systems [37]. The combination of scalability, efficiency, and accuracy makes this system suitable for a wide range of applications, from biosensors [38] and anomaly detection in machinery [39]. It also holds promise for edge computing tasks such as voice and acoustic wave pattern recognition. The use of elastic inputs enables straightforward signal injections by directly exploiting mechanical vibrations, making it particularly well-suited for integrated, real-time sensing and computing applications [4,40].